\documentclass[preprint,superscriptaddress,amsmath,amssymb,prd,aps,showpacs,floatfix,nofootinbib]{revtex4-1}
\usepackage{graphicx}% Include figure filesa
\usepackage{dcolumn}% Aligntable columns on decimal point
\usepackage{bm}
\usepackage{mathrsfs}% bold math
\usepackage{hyperref}% add hypertext capabilities
\usepackage{amsmath}
\usepackage{amssymb}
\usepackage{bm}
\usepackage{color}
\usepackage{float}
\usepackage{dcolumn}
\usepackage{multirow}
\usepackage{changepage}
\hypersetup{pdftex,colorlinks=true,linkcolor=blue,citecolor=blue,menucolor=black,urlcolor=blue,filecolor=blue}

\begin{document}
\title{Theoretical interpretation of the $\Xi(1620)$ and $\Xi(1690)$ resonances seen in $\Xi_c^+ \to \Xi^- \pi^+ \pi^+$ decay}
\date{\today}

\author{Hai-Peng Li}
\affiliation{Department of Physics, Guangxi Normal University, Guilin 541004, China}
\author{Gong-Jie Zhang}
\affiliation{Department of Physics, Guangxi Normal University, Guilin 541004, China}

\author{Wei-Hong Liang}
\affiliation{Department of Physics, Guangxi Normal University, Guilin 541004, China}
\affiliation{Guangxi Key Laboratory of Nuclear Physics and Technology, Guangxi Normal University, Guilin 541004, China}

\author{E. Oset}
\affiliation{Department of Physics, Guangxi Normal University, Guilin 541004, China}
\affiliation{Departamento de F\'{\i}sica Te\'orica and IFIC, Centro Mixto Universidad de
Valencia-CSIC Institutos de Investigaci\'on de Paterna, Aptdo.22085,
46071 Valencia, Spain}

\begin{abstract}
We study the Belle reaction $\Xi_c^+ \to \Xi^- \pi^+ \pi^+$  looking at the mass distribution of $\pi^+ \Xi$,
where clear signals for the $\Xi(1620)$ and $\Xi(1690)$ resonances are seen.
These two resonances are generated dynamically from the interaction in coupled channels of $\pi \Xi, \bar K \Lambda, \bar K \Sigma$
and $\eta \Xi$ within the chiral unitary approach.
Yet, the weak decay process at the quark level,
together with the hadronization to produce pairs of mesons,
does not produce the $\pi \pi \Xi$ final state.
In order to produce this state one must make transitions from the $\bar K \Lambda, \bar K \Sigma$ and $\eta \Xi$ components to $\pi \Xi$,
and this interaction is what produces the resonances.
So, the reaction offers a good test for the molecular picture of these resonances.
Adding the contribution of the $\Xi^*(1530)$ and some background
we are able to get a good reproduction of the mass distribution showing the signatures of the two resonances as found in the experiment.

\end{abstract}

\maketitle

\section{Introduction}
\label{sec:intro}
The $\Xi$ states are still relatively poorly known and only a few of them have been observed \cite{pdg}.
One of them, the $\Xi(1690)$, has a three star status,
and another one, the $\Xi(1620)$, is rated with only one star,
and both of them appear with unknown spin-parity.
The $\Xi(1620)$ is observed in $\Xi(1620) \to \pi \Xi$ decay with large statistical uncertainties \cite{Briefel,bellefon,ross}
and is not found in Ref.~\cite{Hassall}.
From the theoretical point of view the quark models systematically give the first excited state,
after the $\Xi(\frac{1}{2}^+)$, $\Xi(\frac{3}{2}^+)$ ground states,
at about $1750 \, \rm MeV$ \cite{Isgur,Metsz,Riska,Roberts,Xiao} and so does the algebraic model of Ref.~\cite{Bijker}.
One exception is the Skyrme model of Ref.~\cite{Oh},
assuming a soliton-$h$ system,
where two states of $\frac{1}{2}^-$ with $1616\, \rm MeV$ and $1658\, \rm MeV$ are obtained.
The situation with the $\Xi(1620)$ state is similar to that of the $\Lambda(1405)$,
where quark models overestimate the energy.
In both cases, the advent of chiral dynamics,
implemented with coupled channels and unitarity,
came to the rescue and produced a global picture where these states appear naturally \cite{review}.
In particular, in Refs.~\cite{Bennhold,Carmen} the $\Xi(1620)$ and $\Xi(1690)$ appeared both with $J^P=\frac{1}{2}^-$,
coming from the interaction of the coupled channels $\pi \Xi$, $\bar K \Lambda$, $\bar K \Sigma, \eta \Xi$.

The advent of new facilities as Belle, LHCb, BES has added a new dimension to hadron spectroscopy
and new states or old ones with excellent statistics are being observed \cite{Olsen,Karliner,slzhurep}.
In particular weak decays of heavy hadrons have added a new dimension to the information available on dynamically generated resonances,
stemming from the interaction of hadron states \cite{weakrev}.
In that line, it was suggested in Ref.~\cite{Miyahara} to study the $\Xi_c \to \pi \pi \Xi$ reaction
and look for the $\Xi(1620)$ and $\Xi(1690)$ resonances in the $\pi \Xi$ spectrum.
Such reaction has been recently measured at Belle \cite{Belleexp} and excellent signals for these resonances have been observed.
The purpose of the present work is to use this experimental information to further dig into the nature of these resonances.

\section{Generation of the $\Xi(1620)$ and $\Xi(1690)$ states}
\label{sec:generate}

Using the chiral unitary approach, we follow closely the work of Ref.~\cite{Bennhold} and consider the $\pi \Xi$ scattering
with the $\pi \Xi, \bar K \Lambda, \bar K \Sigma, \eta \Xi$ coupled channels.
The $s$-wave scattering amplitudes in matrix form are given by means of the Bethe-Salpeter (BS) equation
\begin{equation}\label{eq:BS}
  T=[1-VG]^{-1} \, V,
\end{equation}
where $V_{ij}$, the transition potential between channels, is obtained from the chiral Lagrangians and given explicitly in Ref.~\cite{Bennhold}.
The $G$ function is the diagonal loop function of the meson-baryon propagators.
Here we differ a bit from Ref.~\cite{Bennhold} and use the cutoff method to regularize the loop function, given by
\begin{equation}\label{eq:Gfunction}
  G_l=\int_{|\vec q \,| < q_{\rm max}} \frac{{\rm d}^3 q}{(2\pi)^3}\, \frac{1}{2\, w_l(\vec q\,)} \, \frac{M_l}{E_l (\vec q\,)} \, \frac{1}{\sqrt{s}-w_l(\vec q\,) -E_l(\vec q\,) +i\epsilon},
\end{equation}
where $w_l(\vec q\,) = \sqrt{m_l^2 +\vec q\,^2}$, $E_l(\vec q\,) = \sqrt{M_l^2 +\vec q\,^2}$,
and $m_l, M_l$ are the meson, baryon masses of the channel $l$.
$G_l$ is regularized cutting with a three momentum $q_{\rm max}$.
This is the only parameter in the theory,
and input from experiment is needed to determine it.
The results from the Belle experiment \cite{Belleexp} provide this information.
Then the BS equation of Eq.~\eqref{eq:BS} is solved, and poles of the $T$ matrix in the second Riemann sheet are searched for.

There exist two poles found in the $T$ amplitude, which are shown in Table~\ref{tab:poles} with different values of the cutoff $q_{\text{max}}$.
\begin{table}[b]
	\caption{Poles of the $T$ matrix with different values of $q_{\text{max}}$. (in MeV)}
\centering
\begin{tabular*}{0.9\textwidth}{@{\extracolsep{\fill}}c c c c c}
\toprule
$q_{\text{max}}$   & $630$ &$700$ &$750$&$770$  \\
\hline
\multirow{2}*{poles}       & $1569.4+125.7i$ &$1563.7 + 106.1i$&$1558.0 + 94.0i$  &$1555.6 + 89.7i$  \\
              & $1687.9 + 0.7i$ &$1681.8 + 1.8i $     &$1674.8 + 2.3i$&$ 1671.5 + 2.4i$  \\
\hline\hline
\end{tabular*}
\label{tab:poles}
\end{table}

The two poles correspond to the $\Xi^*$ resonances generated dynamically from the coupled-channel $\pi \Xi$ interaction,
having the quantum numbers of isospin $I=\frac{1}{2}$ and spin-parity $J^P=\frac{1}{2}^-$.
The first pole locates below the $\bar K \Lambda$ threshold ($\sim 1611 \, \rm MeV$) and has a rather wide width.
While the second pole appears near the $\bar K \Sigma$ threshold ($\sim 1689 \, \rm MeV$), with a narrow width.
We  identify the first pole and the second pole as the $\Xi(1620)$ and $\Xi(1690)$ states, respectively.
According to Table~\ref{tab:poles}, as the cutoff $q_{\text{max}}$ increases from $630$ to $770 \, \rm MeV$,
the positions of the two poles decrease by less than $17\, \rm MeV$.
This means that the masses of the $\Xi(1620)$ and $\Xi(1690)$ states are not very much dependent on $q_{\text{max}}$.
But their widths are more sensitive to the cutoff.

To further understand the molecular components of the $\Xi(1620)$ and $\Xi(1690)$ states,
we calculated the couplings of the two resonances to different channels,
which are showed in Table~\ref{tab:1} when taking $q_{\text{max}}=630\, \rm MeV$.
\begin{table}[bt]
\renewcommand\arraystretch{1.0}
\centering
\caption{Couplings of the two generated states to different channels (with $q_{\text{max}}=630\,{\rm MeV}$). The bold face numbers indicate the pole position of the states.} \label{tab:1}
\begin{tabular*}{0.9\textwidth}{@{\extracolsep{\fill}}ccccc}
\hline\hline
   $ \bm{1569.4+i125.7}$ & $\pi \Xi$          & $\bar K\Lambda$        & $\bar K\Sigma$    & $\eta \Xi$ \\
   \hline
   $g_i$                   & $-2.0-1.6i$  & $1.9+0.9i$  & $0.7+0.5i$  & $-0.1-0.4i$ \\
   $|g_i|^2$               & $6.6$             & $4.5$            & $0.6$            & $0.1$ \\
   \hline\hline
   $\bm{1687.9 +i 0.7}$  & $\pi \Xi$          & $\bar K\Lambda$        & $\bar K\Sigma$    &$\eta \Xi$ \\
   \hline
   $g_i$                   & $0.1-0.1i$       & $0.2+0.1i$      & $-1.2+0.2i$   & $-0.8+0.1i$ \\
   $|g_i|^2$               & $0.01$             & $0.04$            & $1.5$            & $0.6$ \\
  \hline\hline
\end{tabular*}
\end{table}

One can see that the $\Xi(1620)$ state couples mainly to the  $\pi\Xi$ and $\bar{K}\Lambda$ channels,
while the $\Xi(1690)$ state couples strongly to the $\bar{K}\Sigma$ channel. %the opposite
In Ref.~\cite{Bennhold} where dimensional regularization is used to regularize the loop function,
only one excited $\Xi$ state with mass around $1600\, \rm MeV$,
which is identified as the $\Xi(1620)$,
is generated from the $\pi\Xi$ interaction in coupled channels.
However, there was a strong cusp around the $\bar K \Sigma$ threshold that could be identified with the $\Xi(1690)$ resonance,
see Fig.~1 of Ref.~\cite{Bennhold}.
This situation is common in theoretical studies when poles appear close to a threshold.
There is a continuity, with a smooth transition from a very weakly bound state to a slightly unbound or virtual state,
produced by small changes in the regularization of the loops or the strength of the interaction.
For us there is no much difference between the slightly bound and the slightly unbound states,
they correspond to the same dynamics and the experimental signals are also similar.
In the slightly unbound state,
the signal is seen as a sharp peak, typical of a cusp.
Let us mention in passing that some states admitted as resonances, like the $a_0(980)$ correspond to that case.
One can see this in recent high resolution experiments \cite{BESIII2016},
as well as in theoretical studies \cite{LiangXie,Liangdebas}.
Thus, there is no problem to brand the $\Xi(1690)$ also as a resonance
even if it is a border line state between slightly bound and slightly unbound.

\section{The $\Xi_c^+ \to \pi^+ \pi^+ \Xi^-$ reaction}

\begin{figure}[b]
\begin{center}
\includegraphics[scale=0.65]{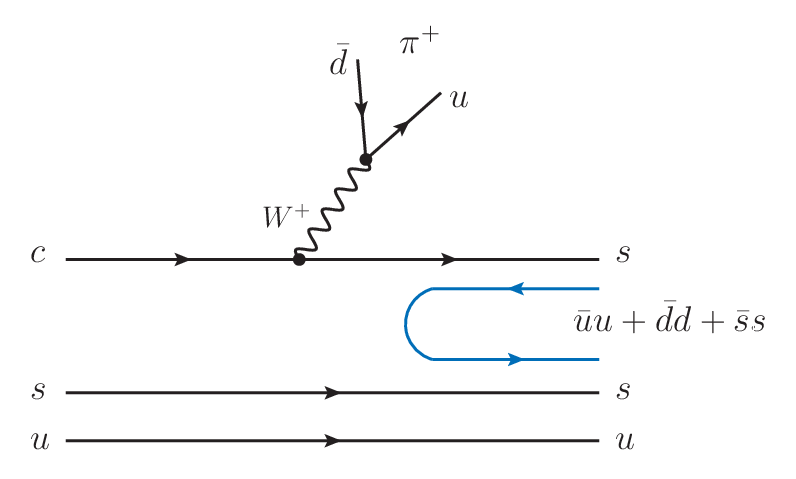}
\end{center}
\vspace{-0.7cm}
\caption{Feynman diagram at quark level for the first step of the $\Xi_{c}^{+}\to\Xi^{-}\pi^{+}\pi^{+}$ decay.}
\label{Fig:1}
\end{figure}

Here we follow closely the formalism of Ref.~\cite{Miyahara}.
At the quark level the reaction to produce the $\Xi(1620)$ and $\Xi(1690)$ proceeds as depicted in Fig.~\ref{Fig:1},
followed by the hadronization of the final quarks,
where a $\bar q q$ state with the quantum numbers of the vacuum is created to produce a meson-baryon pair.

The original $s, u$ quarks are spectators in the weak process and are in a state $\frac{1}{\sqrt{2}}(su-us)$ \cite{Roberts2}.
The weak process is doubly Cabibbo favored and the hadronization must necessarily involve the created $s$ quark after the weak vertex
because in order to have the $\Xi(1620)$ state with negative parity in $\Xi_c^+ \to \pi^+ \,\Xi^0(1620)(\frac{1}{2}^-)$,
the created $s$ quark must have orbital angular momentum $L=1$.
Yet, when the pseudoscalar-baryon coupled channels of section \ref{sec:generate} are created in the hadronization, all quarks are in their ground state,
which forces this $s$ quark to participate in the hadronization process to get deexcited.
Bookkeeping the meson and baryon states coming from the hadronization,
it is found in Ref.~\cite{Miyahara} that the hadronic state coming after this process is given by
\begin{equation}\label{eq:H}
  H= K^- \,\Sigma^+ - \frac{1}{\sqrt{2}}\, \bar K^0 \, \Sigma^0 + \frac{1}{\sqrt{6}} \, \bar K^0 \, \Lambda - \frac{1}{\sqrt{3}} \, \eta \, \Xi^0,
\end{equation}
where the $\eta'$ component, which plays no role in the $\Xi(1620)$ and $\Xi(1690)$ formation, has been neglected.

\begin{figure}[b]
\begin{center}
\includegraphics[scale=0.7]{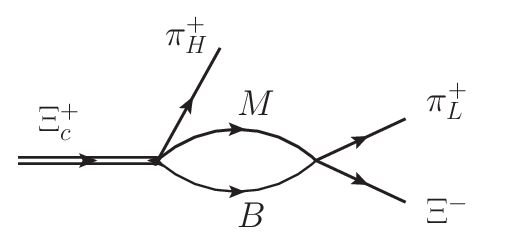}
\end{center}
\vspace{-0.7cm}
\caption{The rescattering mechanism for $\Xi_{c}^{+}\to\pi^{+}\pi^{+}\Xi^{-}$ decay.}
\label{Fig:2}
\end{figure}

It is surprising that the $\pi \,\Xi$ component does not appear in Eq.~\eqref{eq:H},
but it is precisely this feature what stresses the production of a dynamically generated resonance,
because the $\pi \Xi$ final state comes after rescattering of the components of Eq.~\eqref{eq:H} to give $\pi \Xi$ in the final state,
and it is precisely this interaction the one that produces the resonances.
Diagrammatically, this is depicted in Fig.~\ref{Fig:2},
and analytically the decay amplitude corresponding to the mechanism of Fig.~\ref{Fig:2} is given by
\begin{equation}\label{eq:amplitude}
  t=V_P \, \sum_{i=1}^{4} h_i \, G_i (M_{\rm inv}) \, t_{i,\, \pi^+ \, \Xi^-},
\end{equation}
where $V_P$ is a global factor, $t_{i,\, \pi^+ \, \Xi^-}$ are the transition amplitudes from channel $i$ (each state of Eq.~\eqref{eq:H}) to $\pi^+ \, \Xi^-$,
and $h_i$ are the weights of each of these channels in Eq.~\eqref{eq:H},
\begin{equation}\label{eq:hi}
  h_{K^- \Sigma^+}=1;~~~  h_{\bar K^0 \Sigma^0}= -\frac{1}{\sqrt{2}}; ~~~  h_{\bar K^0 \Lambda}= \frac{1}{\sqrt{6}}; ~~~ h_{\eta \Xi^0}= -\frac{1}{\sqrt{3}}.
\end{equation}
Since the $I=\frac{3}{2}$ sector does not contribute to generating the $\Xi(1620)$ and $\Xi(1690)$, we neglect it.
Considering the phase convention of the isospin multiplets $(-\Sigma^+, \Sigma^0, \Sigma^-)$, $(\bar K^0, -K^-)$, $(-\pi^+, \pi^0, \pi^-)$, and $(\Xi^0, -\Xi^-)$, the $t_{i,\, \pi^+ \, \Xi^-}$ amplitudes are obtained from the amplitudes generated in Eq.~\eqref{eq:BS} in isospin basis by means of
\begin{eqnarray}\label{eq:tij}
  t_{K^- \Sigma^+, \, \pi^+ \Xi^-} &=& -\frac{2}{3} \;t_{\bar K \Sigma, \, \pi \Xi}^{I=1/2},  \nonumber\\[2mm]
  t_{\bar K^0 \Sigma^0, \, \pi^+ \Xi^-} &=& \frac{\sqrt{2}}{3} \; t_{\bar K \Sigma, \, \pi \Xi}^{I=1/2},  \nonumber\\[2mm]
  t_{\bar K^0 \Lambda, \, \pi^+ \Xi^-} &=& \sqrt{\frac{2}{3}} \; t_{\bar K \Lambda, \, \pi \Xi}^{I=1/2},  \\[2mm]
  t_{\eta \Xi^0, \, \pi^+ \Xi^-} &=& \sqrt{\frac{2}{3}} \; t_{\eta \Xi, \, \pi \Xi}^{I=1/2}.  \nonumber
\end{eqnarray}

\begin{figure}[t]
\begin{center}
\includegraphics[scale=0.7]{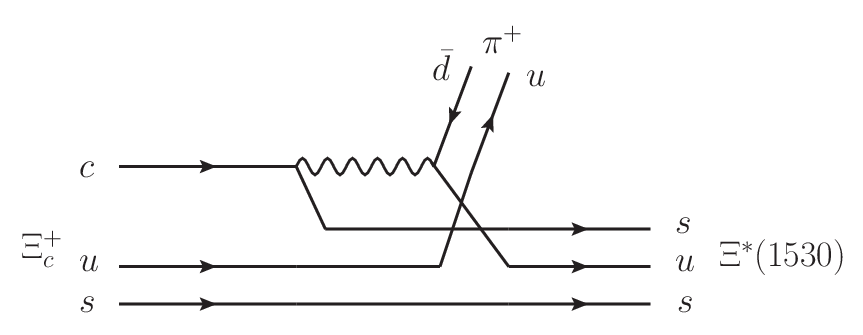}
\end{center}
\vspace{-0.7cm}
\caption{Mechanism for $\Xi^*(1530)(\frac{3}{2}^+)$ production.}
\label{Fig:3}
\end{figure}

In the experiment of Ref.~\cite{Belleexp},
apart from the signals of the $\Xi(1620)$ and $\Xi(1690)$ one can see a clean peak for the $\Xi^*(1530)(\frac{3}{2}^+)$.
It is interesting to see how this is produced.
It cannot be produced with the mechanism of external emission of Fig.~\ref{Fig:1}
because the final wave function is $\frac{1}{\sqrt{2}} s\,(su-us)$,
which is mixed antisymmetric and orthogonal to the fully symmetric $(ssu+sus+uss)$ of the $\Xi^*(1530)(\frac{3}{2}^+)$.
Then one needs internal emission with quark rearrangement as depicted in Fig.~\ref{Fig:3}.

This observation is relevant because the external emission process has bigger strength than internal emission,
but the hadronization also leads to a reduction with respect to a process that requires no hadronization,
as the one in Fig.~\ref{Fig:3}.
As a consequence of this,
one can qualitatively understand that the integrated strengths of the excitation of the $\Xi^*(1530)$ and $\Xi (1620)$ are similar,
as one can observe in the experiment.
We are only concerned here about the $\Xi(1620)$ and $\Xi(1690)$,
but to compare our results with experiment we will account for the contribution of the $\Xi^*(1530)$ empirically.

\section{Results}
\label{sec:results}
The differential width distribution in terms of the $ \pi^+ \Xi^-$ invariant mass is given by
\begin{equation}\label{eq:dGamma}
  \frac{{\rm d} \Gamma}{{\rm d} M_{\rm inv}} = \frac{1}{(2\pi)^3} \, \frac{1}{4\, M^2_{\Xi_c}} \, p_{\pi^+} \, \tilde{p}_{\pi^+} \, |t|^2,
\end{equation}
where $t$ is given by Eq.~\eqref{eq:amplitude} for $\Xi(1620)$ and $\Xi(1690)$.
The amplitudes of Eq.~\eqref{eq:tij} contain the contribution of both resonances.
In Eq.~\eqref{eq:dGamma} $p_{\pi^+}, \, \tilde{p}_{\pi^+}$ are the momentum of the $\pi^+$ produced in the weak process, Fig.~\ref{Fig:1}, %(a),
and the momentum of the $\pi^+$ produced in the final $ \pi^+ \Xi^-$ state, in the $\Xi_c$ and $ \pi^+ \Xi^-$ rest frames, respectively,
\begin{eqnarray*}
  p_{\pi^+} &=& \frac{\lambda^{1/2}(M^2_{\Xi_c},\, m^2_\pi,\, M^2_{\rm inv})}{2\, M_{\Xi_c}},  \\[0.2cm]
  \tilde{p}_{\pi^+} &=& \frac{\lambda^{1/2}(M^2_{\rm inv},\, m^2_\pi,\, M^2_{\Xi^-})}{2\, M_{\rm inv}}.
\end{eqnarray*}
The existence of two $\pi^+$ in the final state of the decay, $\pi^+ \pi^+ \Xi^-$, poses in principle problems in the identification since they are identical particles.
However, in the analysis of the experiment, a most useful strategy is followed,
distinguishing the $\pi_L$ and $\pi_H$ where $\pi_L$ corresponds to the pion with lower momentum and $\pi_H$ to the pion with higher momentum.
It is an interesting kinematical exercise to see that both for $\Xi(1620)$ and $\Xi(1690)$,
as well as $\Xi (1530)$ production, $\pi_H$ corresponds to the pion produced in the weak vertex (Fig.~\ref{Fig:1} and Fig.~\ref{Fig:3}),
while $\pi_L$ corresponds to the one of $ \pi^+ \Xi^-$ decay of the $\Xi(1620)$, $\Xi(1690)$, $\Xi (1530)$ states in the range of invariant masses that we study.
The formula of Eq.~\eqref{eq:dGamma}, with $ \pi^+$ corresponding to $\pi_H$ and $\tilde{p}_{\pi^+}$ to $\pi_L$,
has $M_{\rm inv}$ formed from the $\pi^+_L$ and $\Xi^-$,
consistently with the former discussion.
Hence, a measurement of $\frac{{\rm d} \Gamma}{{\rm d} M_{\rm inv}}$ with $M^2_{\rm inv}=(p_{\pi^+_L}+p_{\Xi^-})^2$ provides the magnitude to be compared to our results.
This magnitude is actually provided in the analysis of Ref.~\cite{Belleexp},
allowing a direct comparison of our results with the data.
Note that this distinction was not done in Ref.~\cite{Miyahara},
where the $\Xi^0 \pi^0$ decay mode was suggested to avoid this problem.

Yet, the data contain also the signal for $\Xi(1530)$ production and some background.
We parametrize them in the following way.
The $\Xi^*(1530)(\frac{3}{2}^+)$ production can be depicted as in Fig.~\ref{Fig:2} substituting the loop function by the $\Xi^*(1530)$ propagator.
But the $\Xi_c(\frac{1}{2}^+) \,\pi^+ \,\Xi(\frac{3}{2}^+)$ vertex requires $p$-wave by angular momentum conservation
and the $\Xi(\frac{3}{2}^+)$ also decays in $p$-wave into $\pi \Xi$.
Thus we must introduce an extra $|t'|^2$ of the type
\begin{equation}\label{eq:tprime}
  |t'|^2 =B^2\, |\vec p_{\pi^+}|^2 \, |\tilde{p}_{\pi^+}|^2 \; \big| \frac{1}{M_{\rm inv}-M_{\Xi(1530)}+ i \Gamma_{\Xi(1530)}/2} \big|^2.
\end{equation}
Obviously this term does not interfere with the $s$-wave production of the $\Xi(1620)$ and $\Xi(1690)$.
We can consider the effect of $\frac{3}{2}^+$ background by adding a term $C^2$ to the propagator square in Eq.~\eqref{eq:tprime}.
In addition we can also add a $\frac{1}{2}^+$ background, which will have $s$-wave in the weak vertex and $p$-wave in the $\pi^+ \Xi^-$ final state.
This term also does not interfere with the other two.
Altogether we can calculate $\frac{{\rm d} \Gamma}{{\rm d} M_{\rm inv}}$ using Eq.~\eqref{eq:dGamma} and substituting
\begin{equation}\label{eq:changet}
  |t|^2 \to |t|^2 + B^2\, |\vec p_{\pi^+}|^2 \, |\tilde{p}_{\pi^+}|^2 \; \left[ \big| \frac{1}{M_{\rm inv}-M_{\Xi(1530)}+ i \Gamma_{\Xi(1530)}/2} \big|^2
  +C^2 \right] + D^2 |\tilde{p}_{\pi^+}|^2.
\end{equation}

As to possible background from $\frac{1}{2}^-$ we do not need to take it into account in the region of the $\Xi(1620)$ and $\Xi(1690)$ resonances.
We have seen that $\frac{1}{2}^-$ comes from rescattering and the chiral amplitudes are good in a reasonable range of energies,
containing much more information than just the poles of the resonances.
There are five parameters in total,
\begin{equation*}
	q_{\text{max}},\quad V_{p}^2,\quad B^2,\quad C^2,\quad D^2,
\end{equation*}
which are used to fit the experimental data from Belle~\cite{Belleexp}.
The parameter $V_P$ regulates the strength of the $\Xi(1620)$
and the parameter $q_{\rm max}$ determines the position and relative strength of the $\Xi(1620)$ to the $\Xi(1690)$.
The other three parameters fit the $\Xi(1530)$ signal and the background.

\begin{figure}[t]
\begin{center}
\includegraphics[scale=0.3]{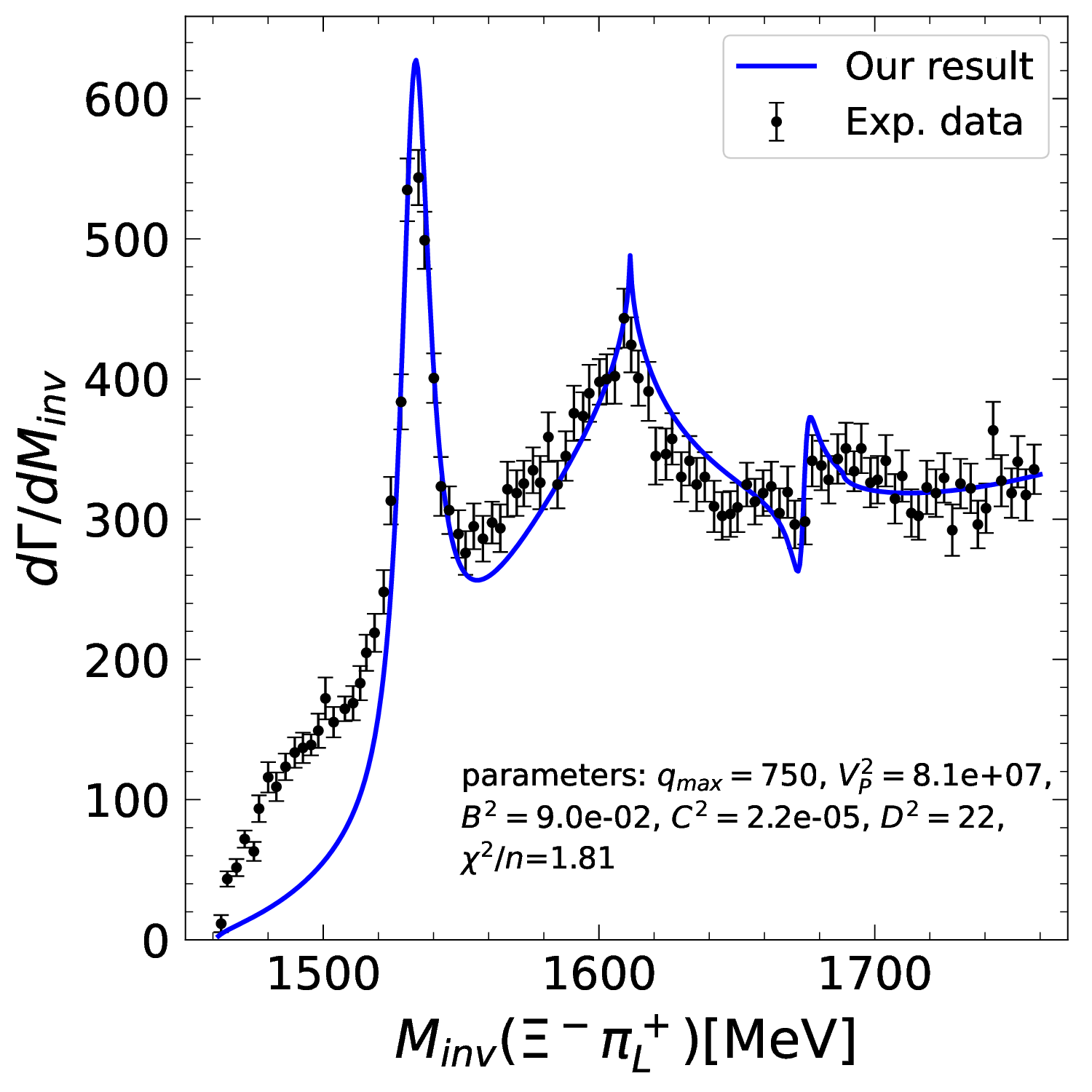}
\end{center}
\vspace{-0.9cm}
\caption{$\Xi^{-}\pi_{L}^{+}$ invariant mass distribution for $\Xi_{c}^{+}\to\pi^{+}_{H}\pi^{+}_{L}\Xi^{-}$ decay.
[parameters: $q_{\text{max}}=750 \,{\rm MeV}$, $V_{P}^2=8.1\times10^7$, $B^2=9\times10^{-2}$, $C^2=2.2\times10^{-5}$, $D^2=22$]}
\label{Fig:4}
\end{figure}

\begin{figure}[t]
\begin{center}
\includegraphics[scale=0.3]{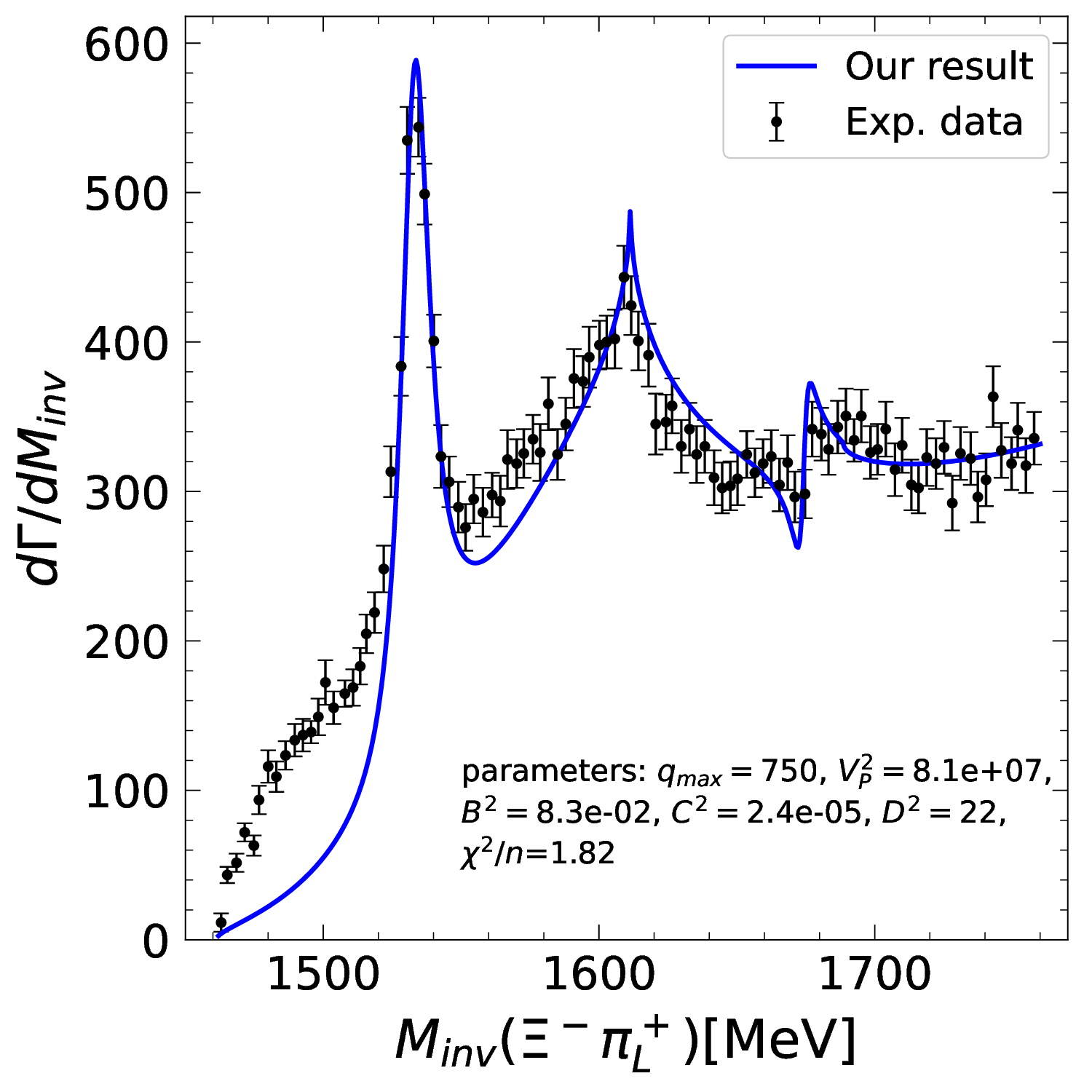}
\end{center}
\vspace{-0.9cm}
\caption{$\Xi^{-}\pi_{L}^{+}$ invariant mass distribution for $\Xi_{c}^{+}\to\pi^{+}_{H}\pi^{+}_{L}\Xi^{-}$ decay.
[parameters: $q_{\text{max}}=750 \,{\rm MeV}$, $V_{P}^2=8.1\times10^7$, $B^2=8.3\times10^{-2}$, $C^2=2.4\times10^{-5}$, $D^2=22$]}
\label{Fig:5}
\end{figure}

\begin{figure}[h]
\begin{center}
\includegraphics[scale=0.3]{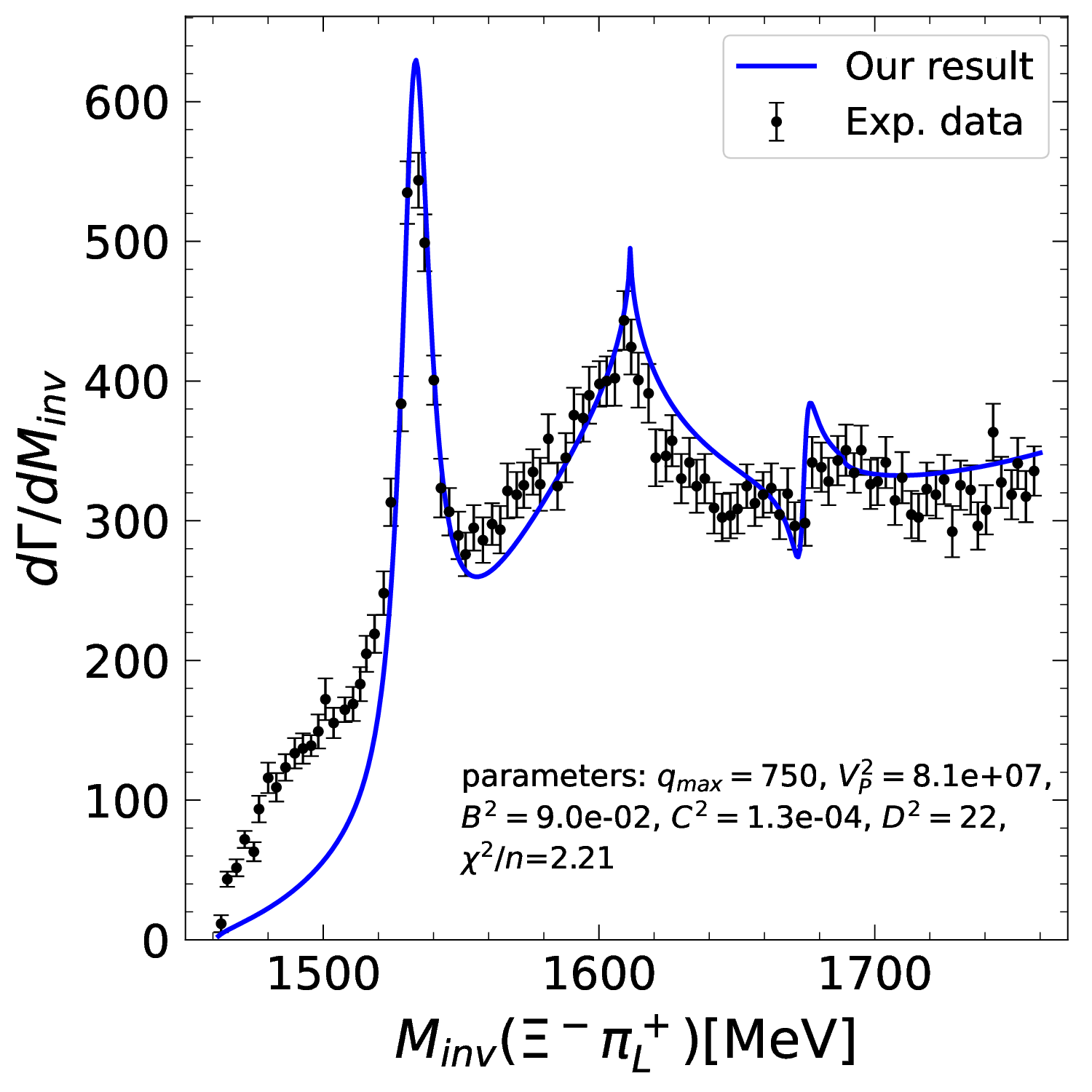}
\end{center}
\vspace{-0.9cm}
\caption{$\Xi^{-}\pi_{L}^{+}$ invariant mass distribution for $\Xi_{c}^{+}\to\pi^{+}_{H}\pi^{+}_{L}\Xi^{-}$ decay.
[parameters: $q_{\text{max}}=750 \,{\rm MeV}$, $V_{P}^2=8.1\times10^7$, $B^2=9\times10^{-2}$, $C^2=1.3\times10^{-4}$, $D^2=22$]}
\label{Fig:6}
\end{figure}

\begin{figure}[h]
\begin{center}
\includegraphics[scale=0.3]{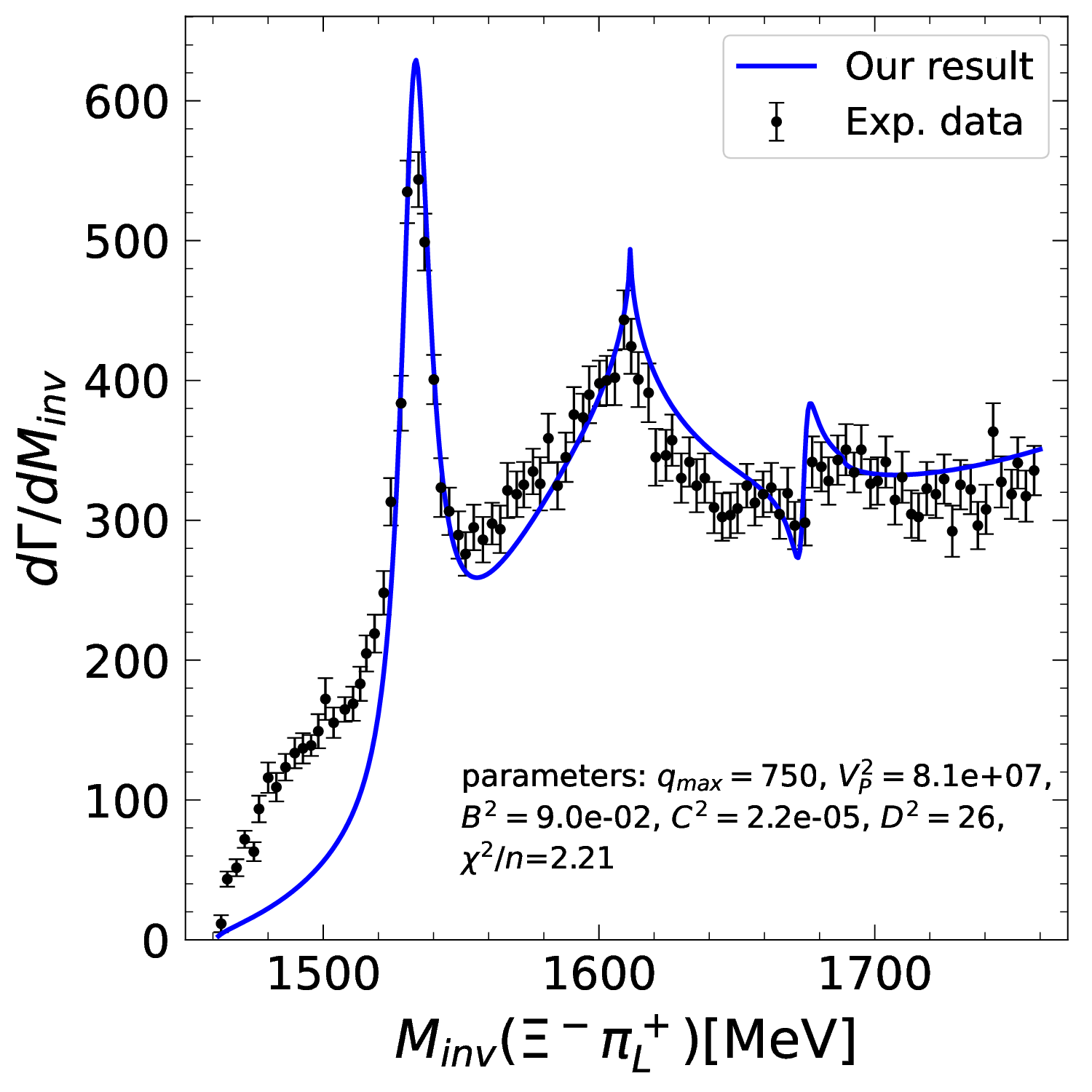}
\end{center}
\vspace{-0.9cm}
\caption{$\Xi^{-}\pi_{L}^{+}$ invariant mass distribution for $\Xi_{c}^{+}\to\pi^{+}_{H}\pi^{+}_{L}\Xi^{-}$ decay.
[parameters: $q_{\text{max}}=750 \,{\rm MeV}$, $V_{P}^2=8.1\times10^7$, $B^2=9.0\times10^{-2}$, $C^2=2.2\times10^{-5}$, $D^2=26$]}
\label{Fig:7}
\end{figure}

\begin{figure}[h]
\begin{center}
\includegraphics[scale=0.3]{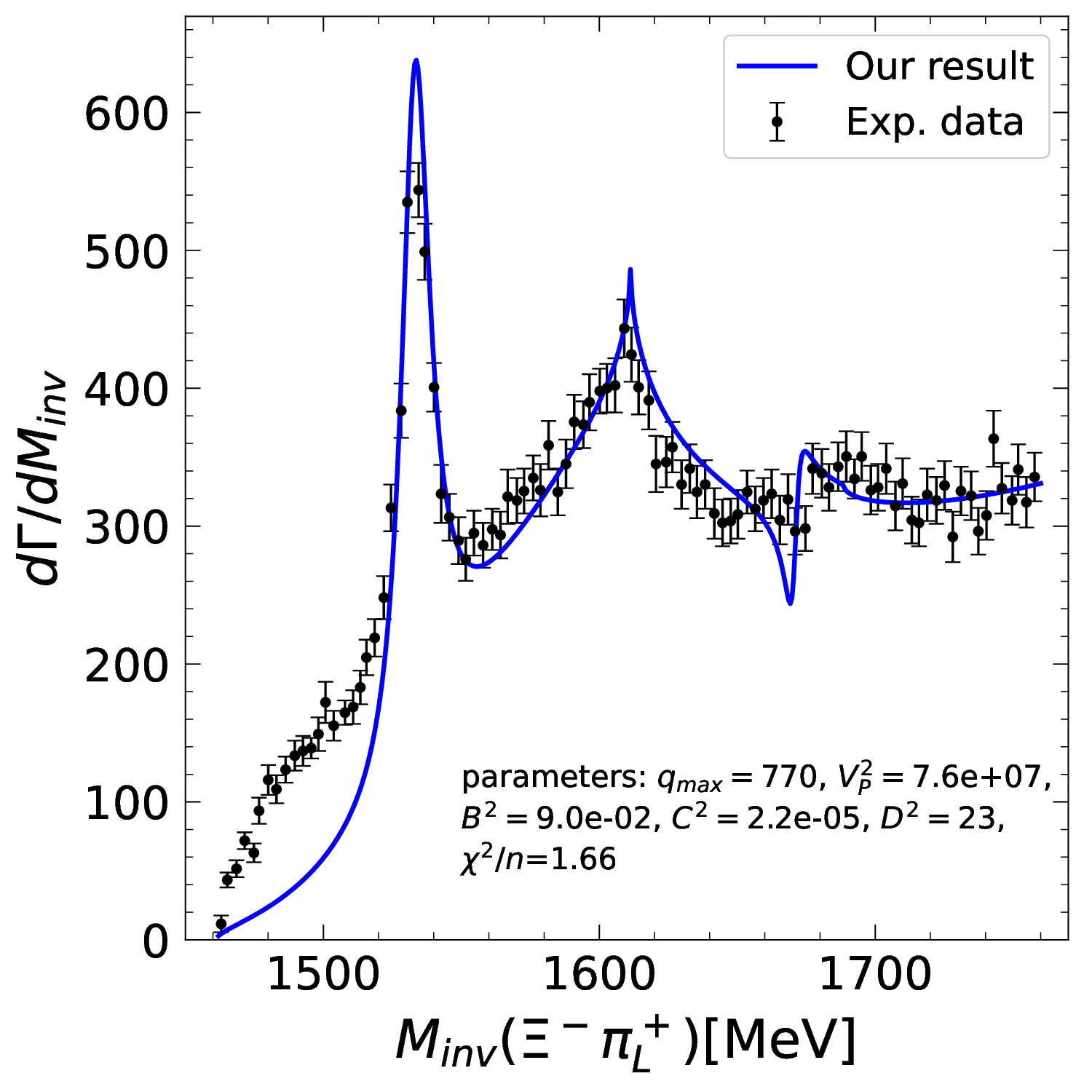}
\end{center}
\vspace{-0.9cm}
\caption{$\Xi^{-}\pi_{L}^{+}$ invariant mass distribution for $\Xi_{c}^{+}\to\pi^{+}_{H}\pi^{+}_{L}\Xi^{-}$ decay.
[parameters: $q_{\text{max}}= 770\, {\rm MeV}$, $V_{P}^2=7.6\times10^7$, $B^2=9\times10^{-2}$, $C^2=2.2\times10^{-5}$, $D^2=23$]}
\label{Fig:8}
\end{figure}

With the aim of seeing more clearly the effect of the individual parameters on the final result, we present the results with different sets of parameters.
In Figs.~\ref{Fig:4}, \ref{Fig:5}, \ref{Fig:6}, \ref{Fig:7}, and \ref{Fig:8},
%
%\clearpage
we show the $\Xi^{-}\pi_{L}^{+}$ invariant mass distribution for $\Xi_{c}^{+}\to\pi^{+}_{H}\pi^{+}_{L}\Xi^{-}$ decay with different sets of parameters,
and the comparison with the experimental data from Belle~\cite{Belleexp},
where the black points with error bar are Belle data and the blue lines are our theoretical results.
The agreement with the data is quite good,
but one finds that the results are inconsistent with experimental data below $1520\, \rm MeV$.

It can be seen that the parameter $B$ mainly determines the strength of the $\Xi(1530)$,
from Figs.~\ref{Fig:4}-\ref{Fig:5}.
Likewise, one can know that the parameters $C^2$ and $D^2$ affect the global strength and the parameter $D^2$ mainly affects the high energy region,
when comparing Figs.~\ref{Fig:4}, \ref{Fig:6} and \ref{Fig:7}.
The position and relative strength of the $\Xi(1620)$ to the $\Xi(1690)$ change
when $q_{\text{max}}$ and $V_{p}^2$ change as shown in Fig.~\ref{Fig:4} and Fig.~\ref{Fig:8}.
We calculate the $\chi^2$ in the invariant mass of $[1530, 1750]\, \rm MeV$, so as to pick out better sets of parameters.
The $\chi^2/n$ is shown in the figure.

In the work we consider that $\Xi(1620)$ and $\Xi(1690)$ are molecular states, with $I(J^{P})=\frac{1}{2}(\frac{1}{2}^{-})$.
The theoretical results are in good agreement with the experimental data in the range of $[1530,1750]\, \rm MeV$,
and support the molecular state picture of the excited  hyperon resonances $\Xi(1620)$ and $\Xi(1690)$.

Although our concern is about the $\Xi(1620)$ and $\Xi(1690)$ resonances,
we make here a short discussion concerning the disagreement in the lower part of the spectrum.
We have tried to overcome this problem by adding some new ingredients:
\begin{enumerate}
\renewcommand{\labelenumi}{\theenumi)}
  \item Extra $\frac{1}{2}^-$ background;
  \item Consideration of the energy dependence of the $\Xi^*(1530)$ width;
  \item Effects of symmetrization of the amplitude. Although, as we mentioned,
  in our resonance amplitudes the $\pi^+_{H}$ and $\pi^+_{L}$ were perfectly identified in the range of energies studied,
  this is not the case for the background.
  Then we performed the calculation
  \begin{equation}\label{eq:add}
    \dfrac{{\rm d}\Gamma}{{\rm d} M^2_{\rm inv}(\pi^+_L \Xi^-)} = \int\dfrac{{\rm d}^2\Gamma}{{\rm d}M^2_{12}\; {\rm d} M^2_{23}} \theta(p_1-p_3)\,{\rm d}M^2_{12}
  +\int\dfrac{{\rm d}^2\Gamma}{{\rm d}M^2_{12}\; {\rm d} M^2_{23}}\theta(p_3-p_1)\, {\rm d} M^2_{23},
  \end{equation}
  using the PDG formula with the order of the particles $\pi^+(1), \pi^-(2), \pi^+(3)$, where
  \begin{equation}
  p_1=\dfrac{\lambda^{1/2}(M^2_{\Xi_c},\, m^2_\pi,\, M^2_{23})}{2\, M_{\Xi_c}}, ~~~~~
  p_3= \dfrac{\lambda^{1/2}(M^2_{\Xi_c},\, m^2_\pi,\, M^2_{12})}{2\, M_{\Xi_c}}.
\end{equation}
\end{enumerate}
Using the different ingredients, we were unable to solve the problem.

Although we are not concerned about this issue here,
the discussion made can serve for future studies trying to find an explanation for this problem.
It would be most interesting to see if an approach like the one of Ref.~\cite{Feijoo,New} going beyond the Weinberg-Tomozawa interaction used here,
has something to say about this issue.

It is interesting to remark that the peak originated by the calculation in ${\rm d}\Gamma / {\rm d} M_{\rm inv}$ in the figures does not exactly correspond to the pole position of the resonance in the second Riemann sheet shown in Table \ref{tab:poles}.
This is a quite common feature in the PDG \cite{pdg} for resonances which are not narrow.
It is also worth mentioning that the $\Xi(1690)$ signal has appeared with a clear interference with the $\Xi(1620)$ and does not have a Breit-Wigner (BW) shape.
It rather looks like ${\rm Im} (iBW)$, with a fast change of sign in the region of the pole.
It is interesting to note that the data with high precision also show a fast jump from one bin to the next in that region.
The molecular picture for the $\Xi(1620)$ and $\Xi(1690)$ that we have used in this work is supported by a large number of papers in the literature in addition to Refs.~\cite{Bennhold, Carmen} mentioned in the Introduction.
The $\Xi(1620)$ can be interpreted as $\bar{K}\Lambda$ molecular state with $I(J^{P})=\frac{1}{2}(\frac{1}{2}^-)$ in Ref.~\cite{Chen:2019uvv},
under the framework of the one-boson-exchange (OBE) model.
This work gives the binding energy for the $\bar{K}\Lambda$ system as $-2.9\, \rm MeV$.
The $\Xi(1620)$ is also assumed to be a $\bar{K}\Lambda$-$\bar{K}\Sigma$ molecular state in Ref.~\cite{Huang:2020taj},
and the decay widths of this state decaying into $\pi\Xi$ and $\pi\pi\Xi$ are calculated
through triangle diagrams in an effective Lagrangian approach,
showing that the $\bar{K}\Lambda$ channel provides about $(50-68)\%$ of the total decay width,
while the $\bar{K}\Sigma$ channel provides the rest.
Similarly, $\Xi(1620)$ is considered a $s$-wave $\bar{K}\Lambda$ or $\bar{K}\Sigma$ bound state,
based on the Bethe-Salpeter equation, in Ref.~\cite{Wang:2019krq},
and the decay widths of $\Xi(1620)\to\pi\Xi$ are $36.94\, \rm MeV$ and $9.35\, \rm MeV$ for the $\bar{K}\Lambda$
and $\bar{K}\Sigma$ bound states, respectively.
However, the $\Xi(1690)$ is not discussed in Refs.~\cite{Chen:2019uvv,Huang:2020taj,Wang:2019krq}.
In our work, the $\Xi(1620)$ is considered as a molecular state with $\pi\Xi$ and $\bar{K}\Lambda$ as its main components,
and the $\Xi(1690)$	is considered as mostly a molecular state of $\bar{K}\Sigma$,
very close to threshold, which could be equally considered as a virtual state, without changing the features of the mass distribution.
An advantage of our work is that we can discuss the $\Xi(1620)$ and $\Xi(1690)$ states at the same time.

The work of Ref.~\cite{Bennhold} is repeated in Ref.~\cite{Sekihara:2015qqa}, by using the chiral unitary approach.
When selecting an appropriate set of subtraction constants, a pole of $1682-0.8i\, \rm MeV$ is found, corresponding to $\Xi(1690)$.
The coupling constants appear to be dominated by $\bar{K}\Sigma$,
which is in line with our results obtained by the cutoff method to regularize the loop function.
More recently, four more works address the nature of these resonances from the molecular perspective.
In Ref. \cite{Nishibuchi},
the $\Xi(1620)$ is obtained using the chiral unitary approach and the $\bar K \Lambda$ scattering length is evaluated.
In Ref. \cite{Alberto},
the same framework is used and the $\Xi(1690)$ is obtained in addition to another resonance,
the $\Xi(2120)$, explaining why the observed widths are so small.
In Ref. \cite{Jujun}, the same approach is used,
the $\Xi(1690)$ is also generated and the related reaction $\Lambda_c^+ \to \Lambda K^+ \bar K^0$ is studied.
In Ref. \cite{Feijoo},
the chiral unitary approach is again used to study the meson-baryon interaction in the $S=-2$ sector
including next to leading order and Born terms in the interaction.
In that work both the $\Xi(1620)$ and $\Xi(1690)$ are generated with similar results and conclusions regarding the nature of the states,
as exposed in the present work.
In our work we dynamically generate the $\Xi(1620)$ and $\Xi(1690)$ states, with only one parameter $q_{\text{max}}$.
We must stress that the only parameter of the chiral unitary approach at this leading order already determines at the same time the position of the two resonances,
their widths and the relative strength of the production rates of the two resonances,
a real challenge for any theory.

\section{Conclusions}
In this work, we have studied $\Xi^{-}\pi^{+}_{L}$ invariant mass distribution for $\Xi^{+}_{c}\to\pi_{H}^{+}\pi^{+}_{L}\Xi^{-}$ decay,
considering the $\Xi(1620)$ and $\Xi(1690)$ as meson-baryon molecular states,
under the framework of the chiral unitary approach.
We can dynamically generate double resonance states simultaneously, $\Xi(1620)$ and $\Xi(1690)$,
from the $s$-wave interaction of $\pi\Xi$ and other coupled channels,
when using the cutoff method to regularize the $G$ functions.
We provide the coupling constants of $\Xi(1620)$ and $\Xi(1690)$ to different channels in Table \ref{tab:1}.
From the coupling constants, it can be known that $\Xi(1620)$ is a molecular state with main components $\pi\Xi$ and $\bar{K}\Lambda$,
and $\Xi(1690)$ is a molecular state with main component $\bar{K}\Sigma$.
The mechanism of $\Xi^{+}_{c}\to\pi^{+}MB$ decay was analyzed in Ref.~\cite{Miyahara}.
We follow closely this work and find that the meson and baryon states from hadronization after the weak process do not contain the final state $\pi^{+}\Xi^{-}$, so it must come from rescattering.
But what might be a problem to describe the reaction becomes actually a strong point to look into the nature of the $\Xi(1620)$ and $\Xi(1690)$ resonances,
which are produced by this final state interaction in the chiral unitary approach.

In order to compare with the experimental data from Belle,
we take into account contributions from $\Xi^*(1530)$ and other backgrounds,
and get the $\Xi^{-}\pi^{+}_{L}$ invariant mass distribution for $\Xi_{c}^{+}\to\pi^{+}_{H}\pi^{+}_{L}\Xi^{-}$ decay.
Results in comparison with experimental data strongly support that the $\Xi(1620)$ and $\Xi(1690)$ are molecular states,
with spin-parity being both $\frac{1}{2}^{-}$.
Other reactions requiring the need of final state interaction to produce these resonances would be most welcome to further support this picture.

\section{ACKNOWLEDGEMENT}
This work is partly supported by the National Natural Science Foundation of China under Grant No. 11975083 and No. 12365019, and by the Central Government Guidance Funds for Local Scientific and Technological Development, China (No. Guike ZY22096024).
This work is also partly supported by the Spanish Ministerio de Economia y Competitividad (MINECO) and European FEDER
funds under Contracts No. FIS2017-84038-C2-1-P B, PID2020-112777GB-I00, and by Generalitat Valenciana under contract
PROMETEO/2020/023.
This project has received funding from the European Union Horizon 2020 research and innovation
programme under the program H2020-INFRAIA-2018-1, grant agreement No. 824093 of the STRONG-2020 project.

\end{document}